\documentclass[prb,article]{revtex4}%
\usepackage{amssymb}
\usepackage{graphicx}
\usepackage{amsmath}%
\usepackage{amsfonts}
\providecommand{\U}[1]{\protect\rule{.1in}{.1in}}
\begin{document}
\title{Mean field exact solutions showing charge density wave crossover at low
fillings in the fractional quantum Hall regime}
\date{July 28, 2006}
\author{Alejandro Cabo$^{*,**}$, Francisco Claro$^{**}$, Alejandro P\'erez$^{**}$ and
Jer\'onimo Maze$^{***}$}
\affiliation{$^{*}$ Grupo de F\'{\i}sica Te\'orica, Instituto de Cibern\'etica,
Matem\'atica y F\'{\i}sica, Calle E, No. 309, Vedado, La Habana, Cuba}
\affiliation{$^{**}$ Facultad de F\'{\i}sica, Pontificia Universidad Cat\'olica de Chile,
Vicu\~na Mackenna 4860, 6904411 Macul, Santiago, Chile}
\affiliation{$^{***}$ Physics Department, Harvard University,USA}

\begin{abstract}
\noindent A general analytical framework for the determination of
the mean field states at arbitrary rational filling factors for
the 2DEG in FQHE regime is given. Its use allows to obtain
analytic expressions for the solutions at filling factors of the
form $\nu=1/q$ for arbitrary odd $q$. The analysis can be
performed
 for two general classes of states characterized by $\gamma=1$ or
$\gamma=\frac{1}{2}$ particles per unit cell. Instead of the
periodic peaks of the Wigner solid solution, the new states show
electron densities forming percolating ridges that may favor an
energy decrease through correlated ring of exchange contributions.
Therefore, we estimate that they can realize mean field versions
of the so called Hall Crystal (HC) states. The obtained analytic
HC solution shows the same crystalline symmetry that the
corresponding WC state in its class $\gamma=1$, but a
qualitatively different charge density distribution.  The energy
dependence of the corresponding HC and WC states on the filling
factor is also evaluated here for the class $\gamma=1/2$. The
results show a crossover between HC state and the Wigner crystal,
close to filling 1/7. Therefore, transitions may occur from one to
the other as the electron density is varied. This result is
consistent with recent experimental findings.

\medskip

\noindent PACS numbers: 73.43.Cd, 73.43.-f

\ \bigskip

\end{abstract}
\maketitle

\section{Introduction}

The quantum Hall effect has been the source of continued interest for already
a quarter of a century.\cite{review} As the samples have become cleaner and
the temperature made lower unexpected structures have been seen, suggesting a
richer physics than originally thought. Although the system was believed to
simply transit from a liquid state to a Wigner Crystal as the electron density
is decreased, recent observations show that new structures develop that may
signal transitions between states of different
symmetry.\cite{chen,csathy,pan,pansto} Numerical calculations also find a
variety of states among which a system may transit, notably at Landau levels
higher than the first.\cite{shiba,reza,hald}

The formation of charge density wave states in the thermodynamic
limit may be studied using the Hartree-Fock
approximation.\cite{yosh,clar0} The formalism allows for the
consideration of states of various symmetries, unidirectional
waves, triangular or square lattices, etc. The criterion normally
used to choose among them is to select the state of lowest energy.
Yet none of such approximate states is the true ground state, so
the effect of correlations is a consideration to have in mind. In
particular, depending on the distribution and relative phases of
the occupied single particle states the correlations may be
enhanced or suppressed. Since experiments show cases in which two
different phases appear it is worthwhile to explore the various
symmetries allowing for self-consistent solutions.

In this work we generalize the formalism given in reference
\cite{cabcla} for the particular $\nu$=$1/3$ filling, to perform
an analytic treatment of the Hartree-Fock problem in the lowest
Landau level (LLL) at arbitrary filling. We also will discuss in
more detail particular solutions at filling factors $\nu=1/q$ for
arbitrary odd $q$. For each value of $q$ two classes of states are
considered and indexed by the number of particles per unit cell
$\gamma=1$ and $\gamma=1/2$. For each one of these two groups, two
types of mean field states are associated. One in which the
electrons build up well localized gaussian-like peaks: the Wigner
Crystal (WC),and another in which the charge density forms
hexagonal ridges increasingly resembling a bee hive as the
electron density is decreased: the Hall crystal(HC) state.
Therefore, the low density limit of this new state differs
qualitatively from the one in the Wigner Crystal
states\cite{ferrari,cabo0,cabcla}. One can then expect that
correlations arising from ring exchange will affect more this
state, since percolating paths are open throughout the
structure.\cite{arov,tao1}. For the particular $\gamma=1/2$ class,
we find a crossover between the WC and HC state, showing that
while the Wigner Crystal (WC) in the class has lower energy at
fillings below 1/7, the new HC state however, has lower energy at
larger fillings. Thus, even within mean field theory both states
are competing as precursors to the ground state.

 Before  ending this section, let us comment on the physical relevance
of the parameter $\gamma$, the number of electron per unit cell.
Its importance is related in one sense,  with the fact that it
determines the symmetry of the mean filed solutions. The value
$\gamma=1$ is associated to the Yoshioka-Lee(YL) WC state which in
the Hartree-Fock framework shows the lowest energy pere particle.
Moreover, within the same class $\gamma=1$ there is a HC state
which shows a bee hive like density, and although having higher
mean field energy than the YL state, one can suspect that it can
receive highly  cohesive correlation energy contributions.
Further, the class $\gamma=1/2$, is the next in the, up to now,
identified relevance, since for these states, it was made clear in
Ref. \cite{clar0}, that  although showing higher mean field
energies than the YL states, its states predict energy gasps for
all the relevant filling factors in FQHE.  Since the HC in this
class is also showing a bee hive like density distribution, the
possibility that the correlation energies can  make them
competitive, a least as alternative ground states in the
experimentally detected phase transitions, is a real one.a The
possible relevance of mean field states having higher than two
particle per unit cell is not excluded, but their consideration
should wait for  obtaining their wave functions. In this sense, we
think that precisely the method outlined in this paper, can be
generalized for obtaining $ \gamma<1/2 $ mean field solutions or
at least a large amount of analytical information on them.

In Sec. 2 a formula for the single particle Fock operator is
presented for the case of an hexagonal lattice. Sec. 3 is devoted
to construct a basis in terms of which the Fock operator matrix
representation reduces to $q$ dimensional blocks, where $q$ is the
denominator of the filling fraction. Section 4 presents the exact
determination of  the new HC mean field state for each of the
classes $\gamma=1,1/2$. They are characterized by having a spacial
region of very low charge density around a high symmetry point in
the unit cell, that acquires a zero if the filling is of the form
$1/q$. The case $q=3$ is discussed in more detail. The evaluation
of the energy dependence on the filling factor for the WC and HC
states showing a crossover, for the particular  class of states
$\gamma=1/2$ , is also done in this section. Finally Sec. 5
presents our conclusions.

\section{The Fock operator}

We consider $N_{e}$ electrons on a plane, in a strong perpendicular magnetic
field. For such a system it was proven long ago that at filling less than one
the only consistent mean field solutions have space fluctuations.\cite{clar2}.
We shall here assume the charge density to form a periodic lattice with a unit
cell containing $\gamma$ electrons, this number being fractional or integer.
Denoting by $\phi$ the flux through such cell in units of the flux quantum
$\phi_{o} = hc/e$, one can readily verify the simple relation
\begin{equation}
\phi\text{\ }\nu=\gamma.\label{fundam}%
\end{equation}
The Fock operator in the LLL may then be written in the form \cite{cabcla}

\begin{widetext}
\begin{equation}
H_{HF}=\sum_Qv(\mathbf{Q})\exp (-\frac{r_o^2\mathbf{Q}^2}4)\, T_{r_o^2 \mathbf{%
n\times Q}},  \label{hamiltonian}
\end{equation}
where
\begin{equation}
v(\mathbf{Q})=2\pi r_o^2\rho (\mathbf{Q})\exp (\frac{r_o^2\mathbf{Q}^2}%
4)\left( \frac{1-\delta _{\mathbf{Q,0}}}{r_o\mid \mathbf{Q\mid }}\exp (-%
\frac{r_o^2\mathbf{Q}^2}4)-\sqrt{\frac \pi 2}I_o(\frac{r_o^2\mathbf{Q}^2}%
4)\right) \frac{e^2}{\varepsilon _or_o}.  \label{poten}
\end{equation}
\end{widetext}

Here $r_{o} =\sqrt{ \hslash c/eB }$ is the magnetic length,
$I_{o}(u)$ a modified Bessel function and $\varepsilon_{o}$ the
background dielectric constant. The operator $ T_{\mathbf{a}}$
displaces the function it acts upon in $-\mathbf{a}$ adding a
magnetic phase factor, as defined in Appendix A. The Fourier
components of the periodic charge density $ \rho(\mathbf{x})$ are
defined as usual,

\begin{equation}
\rho(\mathbf{Q})=\frac1{A_{cell}}\int\mathbf{dx\ }\rho(\mathbf{x}%
)\exp(i\mathbf{Q.x}),\label{chardenfour}%
\end{equation}
where $A_{cell}$ is the unit cell area
\begin{equation}
A_{cell}=\mathbf{n.a}_{1}\mathbf{\times a}_{2}=2\pi r_{o}^{2}\phi,
\end{equation}
$\mathbf{n}$ being a unit vector normal to the plane containing the electrons.
Assuming triangular symmetry, the lattice formed by the electrons is then
invariant under translations in the set of vectors
\begin{align}
\mathbf{R} &  =n_{1}\mathbf{a}_{1}+n_{2}\mathbf{a}_{2},\ \ n_{1},n_{2}%
=0,\pm1,\pm2,...;\label{perlatt}\\
\mathbf{a}_{1} &  =a\ (1,0),\\
\mathbf{a}_{2} &  =a\ (\frac12,\frac{\sqrt{3}}2),\ a=\sqrt{\frac{4\pi\phi
}{\sqrt{3}}}r_{o}.\label{adef}%
\end{align}
The vectors $\mathbf{Q}$ in Eq. (4)span all points in the reciprocal lattice
and are given by
\begin{align}
\mathbf{Q} &  =Q_{1}\mathbf{s}_{1}+Q_{2}\mathbf{s}_{2}\label{reciprocal}\\
Q_{1},Q_{2} &  =0,\pm1,\pm2,...\nonumber\\
\mathbf{s}_{1} &  =- \frac1{\phi r_{o}^{2}}\mathbf{n\times a}_{2},\nonumber\\
\mathbf{s}_{2} &  =\ \ \frac1{\phi r_{o}^{2}}\mathbf{n\times a}_{1}%
,\nonumber\\
\mathbf{s}_{i}.\mathbf{a}_{j} &  =2\pi\ \delta_{ij}.\nonumber
\end{align}
The Fourier components of the density obey the sum rule\cite{clar2}
\begin{equation}
\sum_{Q}^{\prime}|2\pi r_{o}^{2}\rho(\mathbf{Q})|^{2}\exp(\frac{r_{o}%
^{2}\mathbf{Q}^{2}}2)=\nu(1-\nu) .
\end{equation}
where the term $\mathbf{Q}=0$ is omitted from the sum. This relation states
that within the mean field approximation the liquid state of uniform density
($\rho(\mathbf{Q})=0$ all finite $\mathbf{Q}$) in the LLL is only possible at
filling one. At fractional fillings the right hand side is finite and so must
be at least one finite wave-vector Fourier component of the charge density.

\section{Block diagonalization of the Fock operator}

The mean field hamiltonian (\ref{hamiltonian}) describes an electron in a
periodic potential and a perpendicular magnetic field, a case for which many
results are known.\cite{brown} An important property is that the single
particle spectrum in the LLL is arranged in non overlapping bands, each with
the same number of states.\cite{clar4} The number of bands equals the
numerator of the flux per cell, the latter assumed a rational. For filling
$\nu=p/q$, with $p,q$ prime to each other, Eq.(\ref{fundam}) yields a flux
$\phi=\gamma q/p$ per plaquette. Assuming $\gamma$ to be a rational this flux
is then also a rational number.

We consider in what follows two simple cases that illustrate how
different values of the flux $\phi$ are to be treated. One is when
this number is an integer, and another when it is half an integer.
For simplicity we set $\gamma= 1, 1/2$ and $\nu=1/q$, so that
$\phi= q, q/2$, respectively. Our results cover the more general
case $\gamma= p, p/2$ and $\nu= p/q$, giving rise to the same
values of the flux we include in the following discussion. Other
cases may be treated using similar methods to the ones described
below.

\subsection{Integer flux quanta per unit cell}

This case is important since it corresponds to the standard WC, in which each
unit cell captures a full electron charge. The flux traversing a plaquette is
$q$ so that the single electron spectrum will have $q$ bands, one of which is
completely filled and the others empty. Since these bands do not overlap
\cite{clar4} the WC state has thus a gap for all values of q, whether even or
odd. Because an essential feature of experiment at not too low filling
fractions is the different behavior at even and odd values of such quantity,
the WC state is not a good candidate for being the mean field precursor to the
true ground state.

Owing to definition (\ref{reciprocal}) the magnetic translations entering the
Fock operator have the form
\begin{equation}
T_{r_{o}^{2} \mathbf{n\times Q}}=T_{-\frac{Q_{2}}q\mathbf{a}_{1}+\frac{Q_{1}%
}q\mathbf{a}_{2}}.\label{trans1}%
\end{equation}
Since the flux piercing the unit cell is an integral number $q$ of flux
quanta, the set of translation operators $T_{R}$ for all the $\mathbf{R}$
defined in (\ref{perlatt}) commute among themselves, allowing to find common
eigenfunctions to all of them. This is not the case for the translations
(\ref{trans1}) since the original unit cell is partitioned in smaller sectors
if $q$ is greater than 1. The basis we shall construct defines a set of
$q$-dimensional subspaces, which are closed under the action of translations
(\ref{trans1}) for all values of $\mathbf{Q}$. \

A first step in finding the basis is to define a set of eigenfunctions
$\chi_{\mathbf{k}}(\mathbf{x})$ of a translation in the vector $-\mathbf{a}%
_{1}/q$ for each value of the momentum $\mathbf{p}$=$\hslash\mathbf{k}$.
Expressed as linear combinations of the functions $\varphi_{\mathbf{k}%
}(\mathbf{x})$ defined in Appendix A, we write them in the form
\begin{equation}
\chi_{\mathbf{k}}(\mathbf{x})=\sum_{s=-\frac{q-1}2}^{\frac{q-1}2}%
c_{s}(\mathbf{k})T_{-\frac sq\mathbf{a}_{1}}\varphi_{\mathbf{k}}%
(\mathbf{x}),\label{functions}%
\end{equation}
where, for definiteness, we have assumed q to be odd. These
functions must obey the condition
\[
T_{-\frac1q\,\mathbf{a}_{1}}\chi_{\mathbf{k}}(x)=\lambda\,\chi_{\mathbf{k}%
}(\mathbf{x}).
\]
One finds for the eigenvalues $\lambda$ and coefficients $c_{s}$ the set of $q
$ solutions
\begin{align}
\lambda^{(r)}(\mathbf{k}) &  =\exp(i\,\frac{\mathbf{k.a}_{1}}q+i \frac{2\pi
r}q)\\
c_{s}^{r}(\mathbf{k}) &  =\frac1{\sqrt{q}}\exp(-i\,\frac{s \mathbf{k.a}_{1}%
}q-i\frac{2\pi rs}q)\,\nonumber\\
r &  =-\frac{q-1}2,...,\frac{q-1}2.\nonumber
\end{align}
Substituting in (\ref{functions}) yields the $q$ eigenfunctions
\begin{align}
\chi_{\mathbf{k}}^{(r)}(\mathbf{x}) &  =\frac1{\sqrt{q}}\sum_{s=-\frac{q-1}%
2}^{\frac{q-1}2}\exp(-i\,\frac{s \mathbf{k.a}_{1}}q-i\frac{2\pi rs}q)T_{-
\frac{s}q\,\mathbf{a}_{1}}\ \varphi_{\mathbf{k}}(\mathbf{x}),\\
\mathbf{k} &  \equiv\mathbf{k}+n \mathbf{s}_{1}+ m \mathbf{s}_{2}%
.\,\;\qquad\,n,m=0,\,\pm1,\,\pm2,...\label{brillouin}%
\end{align}
The last relation expresses the fact that the states in the new basis are
equivalent upon a shift of $\mathbf{k}$ in any linear combination with integer
coefficients, of the unit cell vectors of the reciprocal lattice corresponding
to the periodicity of the density. The equivalence follows from the following
properties: (a) the functions $\varphi_{\mathbf{k}}(\mathbf{x})$ are
eigenfunctions of any translation $T_{\mathbf{R}}$ for lattice vectors
$\mathbf{R}$ given by (\ref{perlatt}), (b) the operator $T_{\mathbf{R}}$
commutes with all translations entering in the definition of $\chi
_{\mathbf{k}}^{(r)}(\mathbf{x})$, and (c) relation (\ref{phase}) in Appendix
A, stating the equivalence between magnetic translations acting on
$\varphi_{\mathbf{k}}(\mathbf{x})$ and a shift in the momentum labelling these functions.

Let us now inspect the effect of a magnetic translation in $\mathbf{a}_{2}/q$
on the new functions. If such a transformation leaves the $q$-plets invariant,
then the matrix reduction of the Hartree-Fock hamiltonian will follow. One
has,\
\begin{equation}
T_{\frac1q\mathbf{a}_{2}}\chi_{\mathbf{k}}^{(r)}(\mathbf{x})=\frac1{\sqrt{q}%
}\sum_{s=-\frac{q-1}2}^{\frac{q-1}2}\exp(-i\,\frac{s\mathbf{k.a}_{1}}%
q-i\frac{2\pi rs}q)T_{\frac1q\mathbf{a}_{2}}T_{\,-\frac sq\mathbf{a}_{1}%
}\varphi_{\mathbf{k}}(\mathbf{x}).
\end{equation}
After using (\ref{commutation}) for changing the order of the two operators
within the sum, it follows that
\begin{equation}
T_{\frac1q\mathbf{a}_{2}}\chi_{\mathbf{k}}^{(r)}(\mathbf{x})=\exp
(-i\,\frac{\mathbf{k.a}_{2}}q)\chi_{\mathbf{k}}^{([r-1])}(\mathbf{x}%
),\label{new1}%
\end{equation}
where the square bracket defines the number in the set $\{-\frac
{(q-1)}2,...,\frac{(q-1)}2\}$ which is equivalent, modulo $q$, to the integer
in the argument. Thus, a magnetic translation in $\frac1q\mathbf{a}_{2}$ just
turns one function in the $q$-plet into another. Besides the properties
already discussed, the basis can be checked to obey
\begin{align}
T_{\mathbf{a}_{1}}\chi_{\mathbf{k}}^{(r)}(\mathbf{x}) &  =\ \exp
(-i\ \mathbf{k.a}_{1}\,)\chi_{\mathbf{k}}^{(r)}(\mathbf{x}),\\
T_{\mathbf{a}_{2}}\chi_{\mathbf{k}}^{(r)}(\mathbf{x}) &  =\ \exp
(-i\ \mathbf{k.a}_{2}\,)\chi_{\mathbf{k}}^{(r)}(\mathbf{x}),\\
P\chi_{\mathbf{k}}^{(r)}(\mathbf{x}) &  =\chi_{-\mathbf{k}}^{(-r)}%
(\mathbf{x}),
\end{align}
where the parity transformation $P$ is defined as usual, $P\ \chi_{\mathbf{k}%
}^{(r)}(\mathbf{x})=\chi_{\mathbf{k}}^{(r)}(-\mathbf{x}).$

>From the above considerations it follows that the q-dimensional subspace
spanned by the functions $\chi_{\mathbf{k}}^{(r)}(\mathbf{x})$ at fixed values
of $\mathbf{k}$ is left invariant by the action of the operators
(\ref{trans1}) for arbitrary values of $Q_{1}$ and $Q_{2} $. Since the Fock
hamiltonian involves just a sum of such translations, it leaves invariant
these $q$-dimensional subspaces, as well. It is of interest to note that this
is a purely kinematic result which does not depend on the form of the
interaction potential.

In the new basis, the $q^{2}$ matrix elements of the hamiltonian
(\ref{hamiltonian}) can be readily found to have the convenient form
\begin{widetext}
\begin{eqnarray}
h_{\mathbf{k}}^{(r^{\prime },r)} &=&<\chi _{\mathbf{k}}^{(r^{\prime
})}|H_{HF}|\chi _{\mathbf{k}}^{(r)}>  \label{heigen} \\
&=&\sum_Qv(\mathbf{Q})\exp (-\frac{r_o^2\mathbf{Q}^2}4)\exp \left( -i\,%
\mathbf{k.n\times Q}\,\,r_o^2+i \frac{\pi}qQ_2(2r+Q_1)\right)
\delta _{r^{\prime },[r-Q_1]}. \nonumber
\end{eqnarray}
\end{widetext}
The problem has thus been reduced to the self-consistent diagonalization of a
$q$-dimensional matrix for each value of the wave vector $\mathbf{k}$. For a
sample of surface S the degeneracy $D=BS/\phi_{o}$ of the Landau level of the
non interacting problem is then split into $q$ bands that span their range as
$\mathbf{k}$ covers the Brillouin zone, each holding exactly $D/q$
single-particle states.

\subsection{The $\gamma=1/2$ class}

We next turn our attention to the case $\nu=1/q, \gamma=1/2$ for which,
following Eq. (\ref{fundam}), $\phi=q/2$. This case is particularly
interesting because if $q$ is even, say $q=2r$, then $\phi=r$ and the single
particle spectrum has just $r=q/2$ bands. Since $\nu=(1/2)/r$ only the lowest
of these energy bands has occupied states, yet is only half filled. The Fermi
energy is at the center of the band and the state is metallic. By contrast, if
q is odd there are $q$ bands in the spectrum, one of which is completely
filled and the others empty, leaving the Fermi level in a gap. Even and odd
filling fraction denominators thus show qualitatively different behavior, a
metal or an insulator, as experiment demands. This remarkable property makes
this state a reasonable candidate to be the mean field precursor to the true
ground state of the system.

The magnetic translations defining the Fock operator have now the form,
\begin{equation}
T_{r_{o}^{2}\ \mathbf{n\times
Q}}=T_{-\frac{2}{q}Q_{2}\mathbf{a}_{1}+\frac
{2}{q}Q_{1}\mathbf{a}_{2}}. \label{trans2}
\end{equation}
We define a doublet invariant under magnetic translations in the vectors
$\mathbf{a}_{1}$ and $\mathbf{a}_{2},$
\begin{equation}
\varphi_{\mathbf{k}}^{\sigma}(\mathbf{x})=\frac{1}{\sqrt{2}}(\varphi
_{\mathbf{k}}(\mathbf{x})+\frac{\sigma}{\exp(-i\mathbf{a}_{2}.\mathbf{k}%
)}T_{\mathbf{a}_{2}}\varphi_{\mathbf{k}}(\mathbf{x})),\text{ }\sigma=\pm1.
\end{equation}
These functions obey
\begin{align}
T_{\mathbf{a}_{1}}\varphi_{\mathbf{k}}^{\sigma}(\mathbf{x})  & =\exp
(-i\mathbf{a}_{1}.\mathbf{k})\varphi_{\mathbf{k}}^{-\sigma}(\mathbf{x}%
),\label{transgum}\\
T_{\mathbf{a}_{2}}\varphi_{\mathbf{k}}^{\sigma}(\mathbf{x})  & =\sigma
\exp(-i\mathbf{a}_{2}.\mathbf{k})\varphi_{\mathbf{k}}^{\sigma}(\mathbf{x}%
),\nonumber
\end{align}
being turned into each other by translations along the axes, save for a phase
factor. In analogy with the previous subsection we define next a set of
eigenfunctions of the translations $T_{\frac{2}{q}\mathbf{a}_{1}}$, with the
form
\begin{align}
\chi_{\mathbf{k}}^{(r,\sigma)}(\mathbf{x})  & =\frac{1}{\sqrt{q}}%
\sum_{s=-\frac{q-1}{2}}^{\frac{q-1}{2}}\exp(i\,\frac{2\mathbf{k.a}_{1}s}%
{q}\,-i\frac{2\pi rs}{q})T_{\frac{2s}{q}\mathbf{a}_{1}}\ \varphi_{\mathbf{k}%
}^{\sigma}(\mathbf{x}),\label{ji}\\
\mathbf{k}  & \equiv\mathbf{k}+n\text{ }\mathbf{s}_{1}/2+m\text{ }%
\mathbf{s}_{2}/2,\,\;\;\,n,\ m=0,\,\pm1,\,\pm2,...\nonumber
\end{align}
Note that the last line indicates for this case that the equivalence of states
is now under shifts in half the reciprocal lattice unit cell vectors. This is
related to fact that magnetic translations in vectors (6) are non commuting,
so that they have no common eigenfunctions. However, translations in twice the
spatial unit cell vectors, are commuting operations. Similarly as in the
previous subsection, the equivalence follows after considering that, (a) the
functions $\varphi_{\mathbf{k}}(\mathbf{x})$ are eigenfunctions of any
operator $T_{2\mathbf{R}}$ for lattice vectors $\mathbf{R}$ given by
(\ref{perlatt}), (b) $T_{2\mathbf{R}}$ commutes with all the translations
entering in the definition of $\chi_{\mathbf{k}}^{(r,\sigma)}(\mathbf{x})$
through the original functions $\varphi_{\mathbf{k}}(\mathbf{x})$ and (c)
relation (\ref{phase}) in Appendix A, expressing the equivalence between
magnetic translation on $\varphi_{\mathbf{k}}(\mathbf{x})$, with a shift in
the momentum labelling these functions.

Aside from relations (\ref{transgum}) these functions satisfy the following
transformations
\begin{align}
T_{\frac2q\,\mathbf{a}_{1}}\chi_{\mathbf{k}}^{(r,\sigma)}(x)  & =\exp
(i\,\frac{2\mathbf{k.a}_{1}}q + i\frac{2\pi r}q)\,\chi_{\mathbf{k}}%
^{(r,\sigma)}(x),\label{translat1}\\
T_{\frac2q\,\mathbf{a}_{2}}\chi_{\mathbf{k}}^{(r,\sigma)}(x)  & =\exp
(-i\,\frac2q\mathbf{k.a}_{2})\,\chi_{\mathbf{k}}^{([r-2],\sigma)}%
(x).\label{translat2}%
\end{align}
Consider now fixed values of the quantum numbers $(\mathbf{k},\sigma).$
Relations (\ref{translat1}) and (\ref{translat2}) directly show that all
translations (\ref{trans2}) included in the Fock operator (\ref{hamiltonian})
leave invariant the $q$-dimensional subspace spanned by the set {$\chi
_{\mathbf{k}}^{(r,\sigma)}(x)$, all r}.

Using the commutation properties (\ref{commutation}) and relations
(\ref{translat1}) and (\ref{translat2}) the matrix elements of the hamiltonian
(\ref{hamiltonian}) can now be written as \begin{widetext}
\begin{eqnarray}
h_{\mathbf{k}}^{(r^{\prime },r)} &=&<\chi _{\mathbf{k}}^{(r^{\prime },\sigma
)}|H_{HF}|\chi _{\mathbf{k}}^{(r,\sigma )}> \label{heigen2}  \\
&=&\sum_Qv(\mathbf{Q})\exp (-\frac{r_o^2\mathbf{Q}^2}4)\exp \left(
-i\mathbf{k.n\times Q}\,\,r_o^2+i\frac{2\pi Q_2}q(r+ Q_1)\right)
\delta _{r^{\prime},[r+2Q_1]}. \nonumber
\end{eqnarray}
\end{widetext}
Again the mean field problem has been reduced to the diagonalization of a
$q$-dimensional matrix, but now this must be done for each value of the wave
vector $\mathbf{k}$ and the index $\sigma$. An interesting outcome is that the
matrix representing the hamiltonian is identical for the two values of
$\sigma$, so that its $q$ eigenvalues are twice degenerate$.$
   The obtained  block diagonalization of the problem directly
indicates a procedure for obtaining approximate selfconsistent
solutions. One starts  assumm=ing an initial   density function
showing the periodicity in the lattice $\mathbf{R}$ and perform
its Fourier series in terms of the  momenta in the reciprocal.
This furnish the initial Fourier component of the density $\rho
(\mathbf{q})$ lattice cell. Then, defining a 2D uniformly spaces
partition of the sets  of vectors  $\mathbf{q}$ for which the HF
states are inequivalent, the set of Hamiltonian matrices
associated to these momenta values are diagonalized.  Note that
these matrices are fully defined after specifying the initial
Fourier component of the density. Then, the information in this
diagonalization process furnish the values of the eigenfunctions
coefficients. Next, the density to be employed in a new step can
be calculated from the formula
\begin{equation}
\rho(\mathbf{x})=\sum_{\mathbf{k}}\left\vert \sum_{\sigma=\pm 1} \sum_{r=-(q-1)/2}^{(q-1)/2}g_{r}%
^{0}(\mathbf{k})\chi_{\mathbf{k}}^{(r,\sigma)}(\mathbf{x})\;\right\vert
^{2}. \label{density0}
\end{equation}
 The coefficients $g_{r}^{0}(\mathbf{k})$ are the lowest energy
eigenvectors of the reduced Hamiltonian block matrix for each
$\mathbf{k}$ value. Since these matrix blocks do not dependent of
$\sigma$ the eigenvectors are
${g_{r}^{0}(\mathbf{k}),r=(q-1)/2,....-(q-1)/2}$ are also $\sigma$
independent. That is, the same is to be used for both values of
$\sigma$ in the $\gamma=1/2 $ case.
    Further, this new density value is again Fourier expanded in
the reciprocal lattice vectors and the Fourier coefficients
employed to define new Hamiltonian block matrices to be
diagonalized. Then, the described  process should be  developed
iteratively up to the arrival to a convergence. To a situation in
which the density at any new steps closely approaches the one the
previous step.

\section{Analytic form of the  HC states at $\nu=1/q$ for the\ two
$\gamma=1,\frac{1}{2}$ classes}

As discussed at the end of the last section, further progress will
normally require a numerical routine that diagonalizes
self-consistently either (\ref{heigen}) or (\ref{heigen2}), where
the solutions obtained at the end generate the same potential
$v(\mathbf{Q})$ that gave rise to them. There is a special
situations, however, for which further analytic progress is
possible. Because these sates   qualitatively differs from the one
in the usual WC, in a way suggesting the connection with the
states discussed in  \cite{tesanovic} we call it below the Hall
Crystal (HC) state.

Its characterization rests on the following observation. Earlier numerical
work showed that in the special case $\nu=1/q$ there are two pairs of special
self-consistent solutions, one pair for each class of states  $\gamma=1$ or
$\gamma=1/2$ respectively\cite{clar2}. For a given class ($\gamma=1$ or
$\gamma=\frac{1}{2})$ one of the states in its corresponding pair represents
the WC.  This wave-function shows gaussian-like peaks of its density centered
at equivalent symmetry points in the lattice. However, in the second state in
the pair, named before as the Hall Crystal (HC),  the charge density vanishes
at a single point in the unit cell and all its periodic images in the rest of
the crystal. Further inspection in the vicinity of such points showed this
zero to be of order $2\gamma(q-1).$ Since the total particle density is a sum
of the positive definite particle densities of the single particle occupied
states, all associated orbitals must vanish at the special points as the power
$\gamma(q-1).$ This information was extracted form the numerical solutions of
the problems \cite{clar2}. \   Below, we will use this condition to
analytically determine the form the particular  HC  states.

In terms of the basis states defined previously the eigenfunctions of the Fock
operator may be written in the form%

\begin{equation}
\Psi_{\mathbf{k}}^{(b,\sigma)}(\mathbf{x})=\sum_{r=-\frac{q-1}{2}}^{\frac
{q-1}{2}}g_{r}^{b}(\mathbf{k})\chi_{\mathbf{k}}^{(r,\sigma)}(\mathbf{x}%
),\label{wavefn1}%
\end{equation}%
\begin{equation}
\sum_{r=-\frac{q-1}{2}}^{\frac{q-1}{2}}g_{r}^{\ast b}(\mathbf{k})g_{r}%
^{b}(\mathbf{k})=1,\ \label{normaliz}%
\end{equation}
where $b=0,1,...,q-1$ is the band index, and the label $\sigma=\pm1$ is to be
omitted if $\gamma=1$. Because these functions are in the LLL, in our sign
convention ($e=-|e|$) they must be of the form\cite{halrez}
\[
\Psi(z,z^{\ast})=F(z^{\ast})\exp(-\frac{zz^{\ast}}{4r_{o}^{2}}),
\]
where $z=x+iy$, $z^{\ast}=x-iy$, and $F(z^{\ast})$ is an analytic function of
its argument. With no loss of generality we choose one of the special zeroes
to be at the origin. As one approaches this point one expects that
asymptotically $F(z^{\ast})\sim(z\ast)^{s}$, with $s=\gamma(q-1)$. Thus the
function $\Psi$ itself and its first $s-1$ derivatives must vanish at the
origin, giving in all $s$ independent equations to be satisfied. These,
together with the normalization condition (\ref{normaliz}) total $s+1$
equations, sufficient to determine the $q$ coefficients $\left\{  g_{r}%
^{0}(\mathbf{k}),r=-\frac{q-1}{2},...,\frac{q-1}{2}\right\}  $ of the occupied
orbitals. Specifically, for $\gamma=1$ one has $s+1=q$, while for $\gamma=1/2$
one has $2s+1=q$, the factor of two arising from the double valued index
$\sigma$. \ Thus the full structure of the single particle wave functions are
determined after solving the considered set of equations for the coefficients.

Let us examine in more detail the two states  for the class $\gamma=1$ and the
simpler case $q=3.$ The WC has been reported in the literature for this case,
\cite{yosh} but the HC has not. Solving for the coefficients {$g_{-1}%
^{0}(\mathbf{k}),g_{0}^{0}(\mathbf{k}),g_{1}^{0}(\mathbf{k})$} defining the
filled orbitals and substituting them in the equation
\begin{equation}
\rho(\mathbf{x})=\sum_{\mathbf{k}}\left\vert \sum_{r=-1}^{1}g_{r}%
^{0}(\mathbf{k})\chi_{\mathbf{k}}^{(r)}(\mathbf{x})\;\right\vert
^{2},
\label{density1}
\end{equation}
where the first sum runs over all momenta $\mathbf{k}$ in the Brillouin zone,
the particle density may be obtained. The result is shown in Fig. 1 (b), where
the density for the Wigner Crystal solution at the same filling, Fig 1 (a),
has been included for comparison. Notice the presence of sharp hexagonal
channels surrounding low density regions where the density vanishes as the
fourth power of the distance. The fact that the density percolates the
structure much like the wax in a bee hive marks the essential difference
between the HC and WC states. In the latter the charge density is made up
essentially of gaussian functions centered at lattice points as seen in Fig.
1(a). As the filling fraction decreases these gaussian peaks become sharper
and the limit of a classical point-electron WC is approached.

\begin{figure}[h]
\includegraphics[width=4.0in]{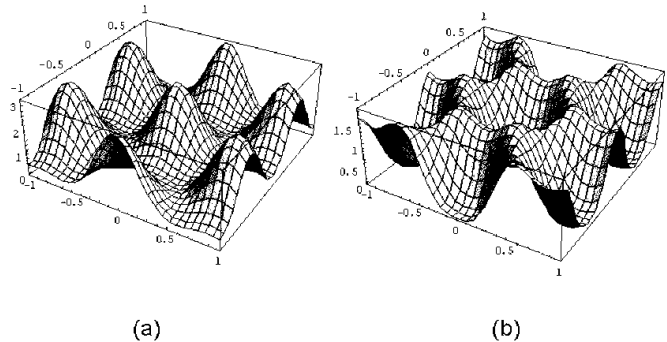}\caption{Particle density for the (a) WC and (b) HC states at
filling $\nu=1/3$ and $\gamma=1$ (one electron per cell). Notice
the percolating hexagonal ridges and the absence of any
gaussian-like peaks in the latter case. }%
\label{density}%
\end{figure}

Further, insertion of the calculated density in Eq. (\ref{chardenfour}) and
use of this result in Eq.(\ref{poten}) allows finding the associated
eigenvalues $\epsilon^{r}$, $r=0,1,2$ by diagonalizing the 3x3 matrix
(\ref{heigen}). Three bands are obtained, that span their range as
$\mathbf{k}$ covers the Brillouin zone.

\begin{figure}[h]
\includegraphics[width=3.0in]{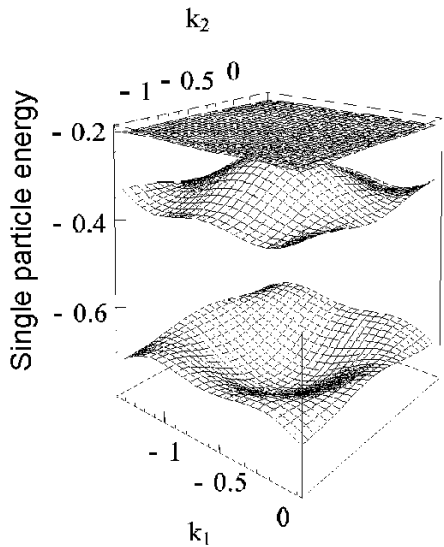}\caption{Dispersion relation for the
three bands into which the lowest Landau level is split by the action of the
Coulomb interaction, for the same case as in Fig. 1. Energy units are
$\frac{e^{2}}{\varepsilon_{o}r_{o}}$.}%
\label{bands}%
\end{figure}

The bands dispersion relations are illustrated in Fig. 2. They are quite
narrow, with the lowest (the filled one) well separated from the rest by a
sizable gap. The same pattern was found for larger values of $q$. The LLL
having been split into $q$ separate bands appears to yield the spectrum
associated with a value of the magnetic field reduced by a factor $1/q$. This
feature is reminiscent of the composite fermion theory, which interprets the
FQHE as the integer QHE of composite fermions in a field reduced by the same
factor.\cite{book,clar3}

The energy per particle of the HC solution we have just discussed is found
through
\begin{equation}
{\Large \epsilon}=\frac{1}{N}\sum_{\mathbf{k}}\frac{{\Large \epsilon}%
^{0}(\mathbf{k})}{2}=-0.362\frac{e^{2}}{\varepsilon_{o}r_{o}},\label{energy}%
\end{equation}
The value obtained is 7\% above the WC solution \ associated to
the $\gamma
$=$1$ class,  for which case $\epsilon=-0.388\frac{e^{2}}{\varepsilon_{o}%
r_{o}}$. The usual minimal energy criterion thus points to the WC as a better
candidate for the ground state and the HC an exited state, a conclusion that
persists for larger values of $q$. Still, how these only approximate mean
field solutions are affected by correlations is largely unknown save for
second order perturbation corrections.

\begin{figure}[h]
\includegraphics[width=2.5in]{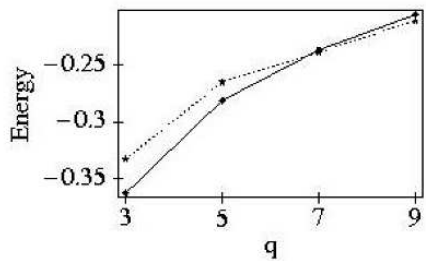}\caption{Energy per particle for
different values of $q=1/\nu$ for the Wigner Crystal (dashed line)
and the Hall Crystal (broken line) both for the class
$\gamma=1/2$. Units are as in Fig. 2. Lines are
drawn to guide the eye.}%
\label{energygraph}%
\end{figure} \

A more interesting case corresponds to the class $\gamma=1/2$ for $q$=$3$. A
similar analytic procedure as employed above gives the energy for both the HC
states. However, It should be noticed  that this special solution was already
discussed in reference \cite{cabcla}. Concretely, the set of three equations
\begin{align}
\sum_{r=-1}^{1}g_{r}^{b}(\mathbf{k})\chi_{\mathbf{k}}^{(r,\sigma)}%
(\mathbf{0})  & =0,\text{ }\sigma=\pm1\\
\sum_{r=-1}^{1}g_{r}^{\ast b}(\mathbf{k})g_{r}^{b}(\mathbf{k})  & =1,
\end{align}
for the coefficients of the functions defining the HC states
through its zero should now be solved. Note that in this case
$\gamma=1/2$  the three equations correspond to the vanishing of
the functions (\ref{wavefn1}) at the origin for the  two values of
\ $\sigma$=$\pm1$ plus the normalization condition
(\ref{normaliz}).

Finally, the dependence of the energy per particle\ on the values
of \ $q$ for the class  $\gamma=1/2$ was evaluated$, $ for the \
WC as well as for \ the HC states. \ The results are shown in Fig.
3. \ Notice that the curves intersect at about filling 1/7,
suggesting that phase transitions may occur between the HC and the
WC states associated to the  class having half \ particle per unit
cell ($\gamma=1/2$), as the magnetic field is varied.
     It should stressed that, although the mean field results for the energy
in the class of states $\gamma=1/2$, depicted in Fig. 3  are
higher than for the ones in the class $\gamma=1$ ( -0.388
$e^2/r_0$ at $\nu=1/3$ for the WC state in this class), these
states could yet enter to compete in having lower energies after
the inclusion of the correlations. Moreover, the fact that recent
experiments detect phase transitions in varying the filling
factors as well as the temperature, could evidence these states as
alternative excited configurations which could become the ground
ones after the temperature and the filling factor are varied.
 These possibilities are consistent with the experimental data
reported in Ref. \cite{chen}.
   The evaluations of the energy per particle, for the WC and HC
states of class  gamma=1/2, were done by numerically solving the
HF problem. The numerical solutions were obtained by using the
same programs  employed in Ref. \cite{ clar0}, which implements an
algorithm  being equivalent to the one sketched at the end of
Section 3. In fact the analytical results presented in Sections 2
and 3, can be interpreted as  obtained in order to perform in an
analytic way a great deal of the numerical steps in the programs
employed for obtaining the results in Ref. \cite{clar0}.

\section{Summary and conclusions}

We have shown that analytical solutions to the Hartree-Fock
problem for a two dimensional electron gas at filling factors
$\nu=p/q$ may be obtained for charge density wave states whose
unit cell include $\gamma$ electrons, with this number fractional
or integral. Our results follow from symmetry considerations and
the construction of a special set of common eigenfunctions to all
magnetic translations in a lattice vector. The spectrum of single
particle states is shown to be organized in multiplets of
dimension $q$ in such a basis, which are left invariant by the
action of the Fock operator. The matrix elements representing the
Fock operator in each multiplet are explicitly determined.

We treat in detail the case $\nu=1/q$ for the two classes holding
one ($\gamma=1$) or a half electron ($\gamma=1$) per plaquette. In
particular for each of such classes,  we analytically determine a
special state whose low density limit is not a Wigner Crystal. We
call this state a Hall Crystal. It is characterized by occupied
orbitals all of which have a zero of order $\gamma(q-1)$ at a
point of high symmetry in each cell. The charge density is made of
percolating hexagonal ridges expected to provide easy paths for
correlated ring exchange that would lower the energy.$\cite{arov}$
For the particular case, $\gamma=1/2$, the mean field energy was
evaluated here as a function of the filling factor and  shows a
cross-over from a regime in which the Wigner Crystal like state
has lower energy, to one in which the Hall Crystal one has lower
energy. In contrast, the same evaluations for the class $\gamma=1$
down to fillings as low as $1/9$, indicates that the Wigner
Crystal like state has the lowest energy within our mean field
approximation. The correlation energy affecting in a different way
both states may alter this ordering, however.

In concluding let us expose some perspectives for the extension of
the work. Let us recall states in the lowest Landau level obeying
periodic boundary conditions must vanish at a number of points
equaling the degeneracy of the non-interacting
system.$\cite{halrez}$ For $\nu=1/q$ this number is $q N_{e}$,
where $N_{e}$ is the total number of electrons in the sample. But,
for example,  the Hall Crystal states at $\gamma=1$ attaches
$(q-1)N_{e}$ of them to fixed high symmetry periodic points in the
electron lattice, leaving $N_{e}$ zeroes whose spatial location
depends on the quantum number $\mathbf{k}$ labelling each occupied
state. Also,  the associated determinant will vanish as the first
power as two particles approach each other. However, the Laughlin
state does so as a power $q$. This feature decreases the direct
Coulomb energy in the Laughlin  states by keeping particles as far
as possible from each other, exhibiting built-in correlations
which mean field solutions lack.
 However, the just described property opens the interesting idea
of constructing new correlated states optimizing short as well as
long range correlations starting form the HC states discussed in
this paper.  For this purpose, we suggest here to substitute the
common factor  of radial functions of the $N_e$ particles which
appear in the HC state  determinant, by Jastrow factors having the
same number of zeroes that factor. In this way,  the so
constructed wave-functions will show $q$ zeros when any two
particle joins. But moreover,  it can be expected that they will
even more optimize the energy,  since they also incorporate the
long range crystalline correlations present in the Hall Crystal
states. We hope the presented results and  ideas for the extension
of the work  will stimulate further research directed to check
whether the corrections to the mean field approach early
introduced  by one of the authors (F.C) could furnish a general
theory for the FQHE.

\begin{acknowledgments}
We acknowledge partial support from Fondecyt, Grants 1060650 and
7020829, the Catholic University of Chile and the Caribbean
Network on Quantum Mechanics, Particles and Fields of the ICTP
Office of External Activities (OEA). One of us (F.C.) would like
to thank the ICTP for hospitality while this work was being done.
\end{acknowledgments}

\appendix

\section{Eigenfunctions of magnetic translations}

It is well known that Bloch-like states may be constructed by placing a seed
function at each lattice point and attaching to it an appropriate phase
factor. In the presence of a magnetic field this procedure may be implemented
using the zero angular momentum eigenfunction
\begin{equation}
\phi(\mathbf{x})=\frac{1}{\sqrt{2\pi}r_{o}}\exp(-\frac{\mathbf{x}^{2}}%
{4r_{o}^{2}})
\end{equation}
and forming the sum\cite{wannier,ferrari,cabo0},
\begin{equation}
\varphi_{\mathbf{k}}(\mathbf{x})=\frac1{N_{\mathbf{k}}}\sum_{\mathbf{\ell}%
}\left(  -1\right)  ^{\ell_{1}\ell_{2}}\exp(i\ \mathbf{k}.\mathbf{\ell
)\ }T\mathbf{_{{\ell}}\ \phi(x),}\label{fi}%
\end{equation}
\[
N_{\mathbf{k}}=\sqrt{N_{\phi_{0}}}\sqrt{\sum_{\mathbf{\ell}}\left(  -1\right)
^{\ell_{1}\ell_{2}}\exp(i\ \mathbf{k}.\mathbf{\ell-}\frac{\mathbf{\ell}^{2}%
}{4r_{0}^{2}}\mathbf{)\ .}}
\]
The summation indices $\ell_{1},\ell_{2}$ run over all integers,
defining a planar lattice $L$ through
$\mathbf{\ell}=\ell_{1}\mathbf{b}_{1}+\ell _{2}\mathbf{b}_{2}$
with the unit cell intercepting one flux quantum, so that
$\mathbf{n.b}_{1}\mathbf{\times b}_{2}=2\pi r_{o}^{2}.$ \
Displacements of the seed function are effected by magnetic
translation operator $T\mathbf{_{a}}$ in the vector $\mathbf{a}$,
whose action on any function $f$ is defined by \cite{avron}
\begin{equation}
T_{\mathbf{a}}\mathbf{\ }f(\mathbf{x})\mathbf{=}\exp\mathbf{(}\frac{ie}{\hbar
c}\mathbf{A({a}).x)\ }f(\mathbf{x-a}).\label{magntra}%
\end{equation}
Here the vector potential is assumed in the axial gauge $\mathbf{A}%
(\mathbf{x})=B(-x_{2},x_{1},0)/2$ and the electron charge $e$ is taken with
its negative sign $(e=-|e|)$. These translations in general do not commute,
\begin{align}
T_{\mathbf{a}_{1}}T_{\mathbf{a}_{2}} &  =\exp(\frac{2ie}{\hbar c}%
\mathbf{A(a}_{1}\mathbf{).a}_{2})T_{\mathbf{a}_{2}}T_{\mathbf{a}_{1}%
,}\label{commutation}\\
&  =\exp(\frac{ie}{\hbar c}\mathbf{A(a}_{1}\mathbf{).a}_{2})T_{\mathbf{a}%
_{1}+\mathbf{a}_{2}}.\nonumber
\end{align}
In the special case of displacements in any vector belonging to L, however,
since the flux trapped by any parallelogram bounded by lattice vectors is an
integral number of flux quanta, all translations commute.

One can easily check that the functions $\varphi_{\mathbf{k}}$ are eigenstates
of translations in any lattice vector, satisfying the eigenvalue equation
\begin{align}
T_{\mathbf{\ell}}\ \varphi_{\mathbf{k}}(\mathbf{x}) &  =\lambda_{\mathbf{k}%
}(\mathbf{\ell})\ \varphi_{\mathbf{k}}(\mathbf{x}),\label{eigen1}\\
\lambda_{\mathbf{k}}(\mathbf{\ell}) &  =\left(  -1\right)  ^{\ell_{1}\ell_{2}%
}\exp(-i\ \mathbf{k}.\mathbf{\ell}).\ \label{eigen2}%
\end{align}
Arranged in a Slater determinant these functions are exact solutions of the
Hartree-Fock problem \cite{cabo0,cabo1,cabo2,tao}. This strong property arises
from the fact that the HF single particle hamiltonian commutes with all
translations leaving $L$ invariant. \cite{cabo0} The functions (\ref{fi}) are
common eigenfunctions of the commuting magnetic translations.\ Moreover, the
set of eigenvalues (\ref{eigen2}) uniquely determines them. Therefore, the HF
hamiltonian associated with the Slater determinant can not change those
eigenvalues and the $\varphi_{\mathbf{k}}$ should be eigenfunctions.

An important property of the basis functions (\ref{fi}) is that an arbitrary
translation is equivalent to a shift in the momentum label, modulo a phase
factor.\cite{ferrari} Operating twice with the translation operator involving
an arbitrary vector $\mathbf{a}$ and a vector in the lattice $\mathbf{\ell}$,
and using Eqs. (\ref{commutation}) and (\ref{eigen1}) one readily gets,
\begin{align}
T_{\mathbf{a}}T_{\mathbf{\ell}}\ \varphi_{\mathbf{k}}(\mathbf{x}) &
=\lambda_{\mathbf{k}}(\mathbf{\ell)}T_{\mathbf{a}}\varphi_{\mathbf{k}%
}(\mathbf{x}),\label{transl1}\\
&  =\exp(\frac{2ie}{\hbar c}\mathbf{A}(\mathbf{a}).\mathbf{\ell})\ T_{\ell
}T_{\mathbf{a}}\ \varphi_{\mathbf{k}}(\mathbf{x}),\nonumber
\end{align}
which can also be written as
\[
T_{\mathbf{\ell}}T_{\mathbf{a}}\ \varphi_{\mathbf{k}}(\mathbf{x}%
)=\lambda_{\mathbf{k}+\frac{2e}{\hbar c}\mathbf{A}(\mathbf{a})}(\mathbf{\ell
})T_{\mathbf{a}}\ \varphi_{\mathbf{k}}(\mathbf{x}).
\]
Then, taking into account that the set of eigenvalues defines uniquely the
wave-functions modulo a phase, it follows that
\begin{equation}
T_{\mathbf{a}}\ \varphi_{\mathbf{k}}(\mathbf{x})=\mathcal{F}_{\mathbf{k}%
}(\mathbf{a})\ \varphi_{\mathbf{k}+\frac{2e}{\hbar c}\mathbf{A}(\mathbf{a}%
)}(\mathbf{x}),\label{phase}%
\end{equation}
where
\begin{equation}
\mathcal{F}_{\mathbf{k}}(\mathbf{a})=\frac{\varphi_{\mathbf{k}}(\mathbf{0}%
)}{\varphi_{\mathbf{k}+\frac{2e}{\hbar c}\mathbf{A}(\mathbf{a})}(\mathbf{a}%
)}.\label{fase1}%
\end{equation}

\end{document}